\documentclass[preprint]{elsarticle}
\usepackage{amsmath, amssymb}

\begin{document}

\title{Comment on the article ``On the Existence of the N-Body Efimov Effect'' by X. P. Wang}

\author{Dmitry K. Gridnev\fnref{fn1}}
\address{FIAS, Ruth-Moufang-Stra{\ss}e 1, D--60438 Frankfurt am Main, Germany}
\fntext[fn1]{On leave from Institute of Physics, St. Petersburg State University, Ulyanovskaya 1, 198504 Russia}
\begin{abstract}
It is shown that the proof of the main theorem in the article ``On the Existence of the N-Body Efimov Effect'' by X. P. Wang, J. Funct. Anal.~209 (2004) pp. 137--161, 
is incorrect. 
\end{abstract}

\maketitle

The rigorous mathematical proof that the Efimov effect exists for three particles has been given in \cite{2}, the discrete spectrum asymptotics was calculated in \cite{3}. 
In \cite{1} the author claims to have extended the proof of the Efimov effect to the case of three clusters. The main result is formulated in Theorem~1.1. Unfortunately, 
there is a mistake in the proof of Theorem~1.1, page 157 of \cite{1}. To show this let us for simplicity denote $C_i = (1 + W_i^{1/2} R_i (\lambda) W_i^{1/2} )^{1/2}$ and 
$D_i = W_i^{1/2} $. For $i = 1,2,3$ these are bounded self--adjoint operators acting on the Hilbert space $\mathcal{H}$. For the definitions of the operators 
$W_i$, $R_b (\lambda), R_i(\lambda), \Pi_0 $ see the notations of \cite{1}. After Eq.~(2.3), page 142 in \cite{1} the author introduces the self--adjoint operator 
$A(\lambda)$ on $\mathcal{H} \otimes \mathcal{H} \otimes \mathcal{H}$, which has the following $3\times 3$ block structure
\begin{equation}\label{a}
 A_{ij}(\lambda) = (1- \delta_{ij}) C_i D_i R_b(\lambda) D_j C_j . 
\end{equation}
The main part of the paper is devoted to the spectral analysis of the operator $F(\lambda)$ (defined on p. 149 in \cite{1}) acting on the same space as 
$A(\lambda)$ and having the following block structure
\begin{equation}\label{f}
 F_{ij}(\lambda) = (1- \delta_{ij}) C_i D_i \Pi_0 R_b(\lambda) D_j C_j . 
\end{equation}
From the inequality $R_b (\lambda) \geq \Pi_0 R_b (\lambda) $ (which is correct) the author concludes that $A(\lambda)$ and $F(\lambda)$ 
are related through
\begin{equation}\label{wrong}
 A(\lambda) - F(\lambda) \geq 0. 
\end{equation}
But (\ref{wrong}) is in general not true. Indeed, even if one takes $C_i, D_i, R_b, \Pi_0$ as linear operators in $\mathcal{H} = \mathbb{R}^n$ (or assuming that $C_i$ 
lie in the appropriate trace ideal) from (\ref{a}) and (\ref{f}) one has 
$\textrm{Tr}\: [A(\lambda) - F(\lambda)] = 0$. And, hence, from (\ref{wrong}) it follows that $A(\lambda) - F(\lambda) = 0$. 
Probably, the confusion came from the fact that the following inequality indeed holds $\mathcal{\tilde A}(\lambda) \geq \mathcal{\tilde F}(\lambda) $, where 
$\mathcal{\tilde A}_{ij}(\lambda) = C_i D_i R_b(\lambda) D_j C_j$ and $\mathcal{\tilde F}_{ij}(\lambda) = C_i D_i \Pi_0 R_b(\lambda) D_j C_j$. 
Bt it is inequality (\ref{wrong}), which is essential for the proof and it is incorrect.

\end{document}